\documentclass{kluwer}
\usepackage{epsfig}

\begin{document}

\begin{article}
\begin{opening}

\title{Search for Spatial Structures at Scales $Z \sim 1$.\\III. The Effect of Lensing on QSO ?}

\author{V.F.Litvin, S.A.Matveef, W.E.Pereira
\footnote{presently at: Dept.of Physics, Michigan Technological University, Houghton, Michigan-49931, USA}}
\institute{Faculty of Physics, St.Petersburg State University,\\Bibliotechnaya Pl.1, Peterhoff,\\St.Petersburg-198904, Russia}

\author{V.V.Orlov}
\institute{Astronomical Institute, St.Petersburg State University,\\ Bibliotechnaya Pl.1, Peterhoff,\\St.Petersburg-198904, Russia}


\begin{abstract}
We carried out a search for peak inhomogeneities in the distribution of 
matter - namely clumps
and voids, within the range $Z \sim 1-3$. We used a new method,
based on the lensing of
quasars by a combination of lenses, belonging to the above 
sought inhomogeneities in the
matter distribution. This work confirms the evidence of the
existence of inhomogeneities found
by us earlier - of a clump ( superattractor N.1), and of a void ( supervoid ). Besides, the
existence of a new gigantic clump ( superattractor N.2 ) was also discovered at $Z \sim 3$.
These clumps could well serve as centers of the Bose-condensation in the early
Universe; in particular - as Anselm's arion condensate, which leads to the formation of
quasiperiodic structures with a period $p \sim 100-200$ Mpc.
\end{abstract}

\end{opening}

\newpage
\section{Introduction}
The structure of the observed part of our Universe-the Metagalaxy-
has been studied so far along the lines of spatial distribution
of galactic clusters and superclusters. These clusters have dimensions ranging from about a few Mpc for the clusters to about
tens of Mpc for the superclusters. It has been assumed that these structures were formed about 10-15 billion years ago. The
more earlier structures could define cosmic objects that were formed before the usual galaxies; quasars, for example. Work
that has been carried out on the statistical analyses of quasar catalogs over the last decade, clearly outlines the quasiperiodic
structures with a characteristic scale around 100 Mpc ( Kunth, 1989; Litvin et al., 1993; Ryabinnikov et al., 1998 ). It is
possible, according to Anselm (1990), that the factor responsible
for the evolution of such anomalies in the matter at the above
mentioned scales, could be a scalar field, which, at a temperature less than the temperature of separation of radiation from a
substance, turns into a condensate. Anomalies in the matter distribution of such a model arise due to a gravitational link between
the matter and the scalar field. However, a strongly unambiguous quasiperiodic structure requires that the condensate layers be
spherically symmetric. This fact was pointed out in the works by Litvin et al. (1994), and it was also shown that a symmetry
like this could be obtained with the presence of a center (or centers) being condensed. The role of such a center could be taken
by the so-called Superattractor (SA) discovered by Litvin et al.(1994).

\par
The hypothesis about the existence of the SA in a given direction on the celestial sphere agrees with the existence of a
maximum density of quasars with absorption lines in that direction (Litvin et al.,1994). Pertaining to this, a case of special
interest is one introduced by Dravskikh \& Dravskikh (1994), on the amplification effect of the observed brightness of quasars
having absorption lines, as opposed to those without absorption lines. This result was confirmed by Van Den Berk et al. (1996).
A possible explanation of this effect in the above works could possibly be the lensing of quasar radiation by massive objects
which make up the absorbing medium. Thus, the search for directions of maximum lensing of quasars over the entire celestial
sphere could become an approach to solving the problem of bringing out the structures in the distribution of gravitating matter
at scales $Z \sim 1$. In the present paper, we propose a method to carry out such a search.

\par
We continue with an analysis, started in the works by Litvin et al.(1997), of Hewitt \& Burbidge's quasar catalog. The analysis
method is based on the search for assumed directions of maximum lensing, and the results obtained are presented in section 2,
with a discussion in section 3. What is important in the present paper, as compared to the one by Litvin et al. (1997), is the
transformation  from equatorial to galactic coordinates. This allows us to most effectively follow random structures that could
arise around the region of the galactic equator, as a result of strong random fluctuations. Details on the probability of the
formation of such structures and statistical analyses of their results are discussed in sections 3 and 4. A more general discussion
and conclusion is then given in section 5.

\section{Method used to search for the lensed matter}
As a parameter characterizing the lensing of quasars, we chose $K_i$, which is the ratio of:
the mean apparent magnitude of the quasar population, falling within the i-th standard cell
(in a circle of radius $R=40^\circ$ on the celestial sphere), $V_i$,
to the mean apparent magnitude over the entire celestial sphere, $V_{4 \pi}$.
\begin{equation}
K_i = {V_i \over V_{4 \pi}}
\end{equation}
\newline
The standard cells are located over the celestial sphere at 150 nodes
of a mesh. The positions of these nodes are chosen by means of a principle, where the celestial sphere quasiuniformly covers these nodes.

\par
This procedure allows us to construct, with the help of a software
called Winsurfer, a topography $K_i$ (l,b) (in the galactic
coordinate system) as a combination of lines of equal height for the 
given red-shift interval $Z_e$, which is defined over QSO emission
lines. Similar to the works by Litvin et al. (1997) we used four 
red-shift intervals, $Z_e \epsilon [0.5 ; 2.5]$;
$Z_e \epsilon [1 ; 3]$; $Z_e \epsilon [1.5 ; 3.5]$; 
$Z_e \epsilon [2 ; 4.5]$. Fig.1 shows the corrresponding topographies
for four populations of quasars from the Hewitt \& Burbidge 
catalog (1993). The main characteristic on all four topographies is 
the presence of well-defined maxima amd minima.

\section{Discussion of results obtained by the above method}
Let us try to discuss a possible interpretation of the presence of 
these ``spots'' (i.e. regions, where the parameter,$K_i$ is 
extremal).

\subsection{\underline{\bf ``Spots'' with a minimum value for $K_i$}}
One observes two such spots on the topographies for the four QSO
populations, at $l_1 = (250^\circ \pm 20^\circ)$;
$b_1 = (-5^\circ \pm 20^\circ)$ and at
$l_2 = (120^\circ \pm 20^\circ)$; $b_2 = (5^\circ \pm 20^\circ)$
respectively. Let us consider each ``spot'' separately.

\subsubsection{\bf{``Spot'' $\bf l_1 = (250^\circ \pm 20^\circ)$; $\bf b_1 = (-5^\circ \pm 20^\circ)$}}
The existence of the region where $K_i$ is minimum in the given
direction may be related to the lensing of quasars, since it is in
this direction that Litvin et al.(1994) had earlier discovered the
maximum density of quasars with absorption lines, and this fact
supports the amplification effect of lensing according to Dravskikh
\& Dravskikh. Besides, in this direction, an anomalous behaviour of
the slope of the straight line on the Hubble diagram also favors the 
lensing hypothesis. This was explained by Litvin et al.(1994) with 
the Superattractor (SA) hypothesis according to which, an increase 
in the average brightness of quasars in this direction may be due to 
an increased probability of quasar lensing by matter from the SA. On 
the other hand, according to Baryshev and Yezova (1997), in the case 
of strong lensing, under a fractal distribution of matter along the 
line of sight, the most probable position of the lens would be, 
either closer to the observer or 
to the source. If
the above mentioned fact
materializes in our case, then a maximum lensing effect will be seen 
for an interval $Z_e$ closer to the approximated position of the SA 
($Z_{SA} = 2.13 \pm 0.60$ according to Litvin et al.1997),i.e. for 
the interval $Z_e \epsilon [1 ; 3]$. And indeed, in Fig.1, we notice 
that the region of minimum $K_i$ at $l_1 = (250^\circ \pm 20^\circ)$;
$b_1 = (-5^\circ \pm 20^\circ)$ is most brightly expressed within the
interval $Z_e \epsilon [1 ; 3]$.

\subsubsection{\bf{``Spot'' $\bf l_2 = (120^\circ \pm 20^\circ)$; $\bf b_2 = (5^\circ \pm 20^\circ)$}}
This spot also appears on the topographies for all four sets of 
quasars, but the effect is maximum for the interval 
$Z_e \epsilon [1.5 ; 3.5]$ and it reduces with a decrease in $Z_e$. 
Thanks to an increased probability of strong lensing of quasars in a 
given direction, we shall follow the above suggested interpretation 
for the region of minimum $K_i$ as being the direction of maximum 
amplification of mean quasar brightness. Let us therefore consider 
the behaviour of the topography of the relative density of absorbers 
in this region, for the interval $Z_e \epsilon [1.5 ; 3.5]$.
The topography corresponding to this region, which was constructed by
us according to the method proposed by Litvin et al.(1997),in 
galactic coordinates, is presented in Fig.2A ( whereas for 
comparison, Fig.2B has the real topography for the lensing parameter 
$K_i$ within the interval $Z_e \epsilon [1.5 ; 3.5]$) The presence of
a spot where the value of the relative density of absorbers is 
maximum $l = (135^\circ \pm 20^\circ)$; $b = (10^\circ \pm 20^\circ)$
in Fig.2A confirms the above hypothesis of the connection between 
the discussed spot $l_2 = (120^\circ \pm 20^\circ)$; 
$b_2 = (5^\circ \pm 20^\circ)$ in the topography in Fig.1 and the 
efect of lensing.

\par
For the given spot, we shall not be discussing results on how the 
topography of the slope of the straight line on Hubble's diagram 
behaves in galactic coordinates. This is because the method used in 
defining the slope is very sensitive to random fluctuations which 
arise due to minor changes in the number of quasars falling within a 
particular cell. The effect becomes even stronger with an increase in
$Z_e$, since an increase in $Z_e$ means a decrease in the total 
amount of data in Hewitt \& Burbidge's catalog (1993). This way, 
wherever there is a discussion on the topography of the slope of the 
straight line on Hubble's diagram we use only a qualitative 
comparison with results from the works by Litvin et al.(1997), since 
in those works we had used equatorial coordinates but had not 
considered the region around the North Pole due to strong distortions
in the Mercatorial projections.

\subsection{\underline{\bf ``Spot'' with a maximum value for $K_i$}}
The position of this spot on the celestial sphere 
$l_3 = (60^\circ \pm 20^\circ)$; $b_3 = (0^\circ \pm 20^\circ)$ 
coincides, within a margin of error, with an earlier discovered, but 
uninterpreted region (Litvin et al.,1994; Litvin et al.,1997) where 
the slope of the straight line on Hubble's diagram is maximum. In 
both cases, the spots were observed only within the interval 
$Z_e \epsilon [2 ; 4.5]$ unlike the spots of minimum $K_i$ 
considered earlier.

\par
As a cause for the appearance of this spot one could assume the 
existence of a gigantic Supervoid in the region $Z_e \sim 3$.

\par
Let us consider, for simplicity, a spherical void. Quasars, located 
at the farther (relative to the observer) boundary of the void, are 
drawn into motion by an unbalanced gravitational attraction of mass 
outside the void region. As a result, the observed red-shift of such 
quasars increases. On the other hand, this effect leads to a decrease
in the red-shift of quasars located at the nearer boundary of the 
sphere. Because of this we observe an increase in the slope of the 
line on Hubble's diagram , which leads to the formation of a spot of 
maximum slope in the direction of the void.

\par
In order to explain, within the bounds of the present hypothesis, 
the formation or occurence of a spot with a maximum mean stellar 
value of quasars on the topography in Fig.1D, it is necessary to 
take into account the distribution of quasars as a function of $Z_e$ 
in Hewitt \& Burbidge's catalog (1993). The number of quasars in a 
unit interval around $Z_e \sim 4.5$ is negligibly small when 
compared to their number around $Z_e \sim 2$.
Thus the effect by which the red-shift of quasars around $Z_e \sim 2$
decreases, can be considered a dominant factor. At the same time, 
quasars having a $Z_e > 2$, but lying around the boundary $Z_e = 2$, 
could "fall out" from the interval $Z_e = [2 ; 4.5]$, these being 
the brightest for the given interval. As a result, we observe an 
overall increase in the mean apparent stellar magnitude within the 
interval $Z_e = [2 ; 4.5]$ in the direction of the void ( spot of 
maximum $K_i$ in Fig.1D ). It should be noted that in the absence of 
the void within other intervals ( Figs.1:A,B,C) the above effect is 
not observed.

\newpage

\section{A statistical analysis of the data obtained}
A detailed analysis of the topograms, about which we had spoken in 
the previous section, shows, that all extremal spots lie close to 
the plane of the galactic equator. This carries a definitive 
scepticism in relation to the proposed interpretation of the results 
obtained by us. Poor statistics in the region near the galactic 
equator leads to an increased influence of random fluctuations in 
the data analysis. In this way, the probability of occurrence of 
random structures, analogous to the spots in Figs.1 and 2, increases.
Moreover, a non-uniform choice of objects from the catalog for 
different areas of the celestial sphere can lead to a selection. In 
order to evaluate the influence of these factors, we carried out an 
analysis of the Hewitt \& Burbidge (1993) catalog using two 
different methods:

\par
$\bf {1.)}$ An estimate was carried out of the probability of random occurrence of ``extremal'' spots on the topograms in Figs.1 and 2.

\par
Similar to section 2, we used the method of constructing topographies
for $K_i$ by taking data from the same catalog. The role of standard 
cells was taken over by N identical domains, e.g. circles of radius 
R, whose centers form a mesh on the celestial sphere. ( For the sake 
of convenience, we used a mesh that was rectangular in the 
Mercatorial projection of the celestial sphere).
The size and the corresponding number of domains were defined by the
necessary condition that $N_{obj} \geq 5$ objects per standard cell.
That way, after creating the topography of the celestial sphere, we 
obtained a set of domains, wherein each of which, the lensing 
parameter $K_i$ was defined.

\par
Now, let the number of such domains be equal to {\sl n}; ( where, 
for a
spot of minimum $K_i$, the relation $K_i\leq K_{cr}$ is true; or
where, for a spot of maximum $K_i$, the relation $K_i\geq K_{cr}$ is
true; $K_{cr}$ being some critical cut-off parameter on the 
topogram ).

\par
Also, let the lensing topogram have a spot made up of {\sl m} 
domains, which satisfy $K_i\leq K_{cr}$ ( or $K_i\geq K_{cr}$, 
in the case of a maximum ). By ``spot'' we imply a set of domains in 
a rectangular mesh, where the center of each new element differs 
from the previous
element by exactly one step of the mesh.

\par
In order to estimate the probability of the random formation of the 
given spot in any part of the topogram, we carried out the following 
procedure: \\
{\sl n} domains with $K_i\leq K_{cr}$ ( or $K_i\geq K_{cr}$, 
in the case of a maximum) were randomly thrown over a sphere and the 
resulting topogram was then scanned to check for at least a single 
case in which the spot would occur, within {\sl m} domains 
satisfying $K_i\leq K_{cr}$ ( or $K_i\geq K_{cr}$, in the case of a 
maximum).

\par
The next step was to calculate {\sl q} such occurrences for {\sl Q} 
throws or iterations. \\
Finally, the ratio $p = q / Q$ gives us the required probability 
value.

\par
Results of our calculations for spots with minimum and maximum values
of the parameter $K_i$ (Figs.1:A,B,C,D) are presented in 
Tables I.a,b,c,d; II.a,b,c,d and III.

\par
Similarly, we can estimate the probability of a random occurrence of 
extremal spots on the topograms of the relative density of absorbers 
( see Fig.2A and also Litvin et al.,1997). These estimations are 
presented in Table IV.

\par
Calculations (Table I.) show that the occurrence of two of the spots
of minimum $K_i$ on the lensing topography (the first one in the
direction $l_1 = (250^\circ \pm 20^\circ)$; 
$b_1 = (-5^\circ \pm 20^\circ)$ 
within the intervals $Z_e \epsilon [0.5 ; 2.5]$;
$Z_e \epsilon [1 ; 3]$; and $Z_e \epsilon [2 ; 4.5]$; and the second
one in the direction  $l_2 = (120^\circ \pm 20^\circ)$; 
$b_2 = (5^\circ \pm 20^\circ)$ within the interval 
$Z_e \epsilon [1.5 ; 3.5]$)
can be considered non-random. However, while doing so, 
it is understood that effects of a selection cannot be omitted 
(see the corresponding estimate below). The same can be said about
spots of maximum $K_i$ on the lensing topogram within the interval
$Z_e \epsilon [2 ; 4.5]$ ( Fig.1D and Table III) and on the topogram
of the relative density of absorbers within the interval
$Z_e \epsilon [1.5 ; 3.5]$ (Fig.2A and Table IV).

\par
$\bf {2.)}$ An estimate was also carried out of the probability of
occurrence of extremal spots on the topogram of Fig.1 as a result of
possible quasar selection effects from the catalog by Hewitt \&
Burbidge (1993).

\par
In this method, we carried out a reapproximation ( ``mixing'') of 
quasar parameters (namely apparent stellar magnitude and redshift), 
but their coordinates remained fixed.
At the same time, using the standard procedure (see section 2.), we 
defined a set of
$N_{knots}$ values for the lensing parameter $K_i$ (where, 
$N_{knots}$-is the number of standard cells, chosen by us, and 
defined by the radius of the cells, in particular, by the condition 
that a sphere entirely covers the cells with a minimum excess 
covering). \\
For a given p-th set of cells we defined a maximum $K_{p}{}^{max}$ 
and a minimum
$K_{p}{}^{min}$ for the value of the lensing parameter $K_i$. \\
We then calculated $n_1$ sets for which the following condition was 
satisfied: \\
$(K_{p}{}^{max} - K_{p}{}^{min}) \geq (K^{max} - K^{min})$, \\
where $K^{max}$ and $K^{min}$ - corresponding maximum and minimum 
values for the lensing parameter, defined for the catalog by Hewitt 
\& Burbidge (1993) without a ``mixing''.

\newpage
The ratio, $p_1 = n_1 / Q_1$ then gives us the estimate of the 
probability of the  formation of spots on the topograms (Figs.1 \& 2)
taking into account the effects of a selection. The results of these 
calculations within four $Z_e$ intervals and for three
values of the radius R of a standard cell, are presented in 
Table V.a,b,c,d.

\par
An analogous estimate of the probability for the topography of 
absorbers in Fig.2A is presented in Table VI.

\par
Results from Tables V \& VI support the fact that the probability of 
occurrence of extremal spots due to closed selection effects are very
small. This, in conjunction with the results obtained above of an 
estimate of the probability of random occurence of these spots (see
above for discussion of Tables I-IV), leads us to conclude that the 
role of random fluctuation
effects is very small, and the formation of anomalies on the 
topographs (Figs.1 \& 2) can be interpreted as a physical effect, 
according to section 2.

\par
In conclusion, it is interesting to note a certain characteristic in 
the behaviour of the spot at minimum $K_i$ as a function of $Z_e$, in
the direction  $l_1 = (250^\circ \pm 20^\circ)$; 
$b_1 = (-5^\circ \pm 20^\circ)$. For this direction, the
lensing effect is maximum within the interval $Z_e \epsilon [1 ; 3]$.
On increasing or decreasing $Z_e$ within the intervals 
$Z_e \epsilon [0.5 ; 2.5]$ and $Z_e \epsilon [1.5 ; 3.5]$ the effect 
weakens (within $Z_e \epsilon [1.5 ; 3.5]$ the effect is almost 
statistically negligible-see Table Ic). However, within 
$Z_e \epsilon [2.5 ; 4.5]$
the effect is stronger and an occurrence by chance of the 
corresponding spot is highly improbable (Table Id.). At first sight, 
such a ``strange'' behaviour of the anomaly in the direction 
$l_1 = (250^\circ \pm 20^\circ)$; $b_1 = (-5^\circ \pm 20^\circ)$ can
be connected to different ``regimes'' of lensing of radiation from 
near and far (relative to the SA) quasars by components of matter 
from the SA.
For quasars located nearer to the SA, in $Z_e \epsilon [1 ; 3]$ 
( i.e. in the region of maximum concentration of lenses) there exists
an increased probability of strong lensing effects as a result of a 
strong probability that the observer could fall in the region of a 
conical caustic (Baryshev \& Yezova, 1997). This was already 
mentioned in section 2.
On increasing the relative lens-quasar distance, the probability of 
strong lensing drops, but the probability of multiple weak-lensing 
increases ( if the distance to the lens is greater than half the 
distance to the quasar). In this case, the maximum probability of 
lensing occurs, if the lens is exactly half-way between the observer 
and source. Hence, when the
lenses are concentrated in the region around $Z \sim 2$, one should 
expect an amplified
lensing effect of quasars at $Z_e \sim 4$, i.e. in our case, within 
the interval $Z_e \epsilon [2 ; 4.5]$.

\newpage
\section{Discussion}
An analysis of the results obtained with the help of the method 
proposed in this
paper,of finding the direction of maximum lensing of quasars, is
indeed a continuation for the search of large-scale inhomogeneities 
in
the distribution of matter at scales $Z \sim 1$. The
existence of one such inhomogeneity around $Z \sim 2$ - the SA, 
which was
found by us earlier while analysing the spatial distribution of 
absorbers, confirms
the presence of a maximum in the mean observed brightness of quasars 
in the
direction of the SA. Yet another maximum
in the mean observed quasar brightness was discovered in the region 
$Z \sim 3$ in
a direction (equatorial coordinates) that we had not searched.
This anomaly, as in the previous case, corresponds to a maximum in 
the density of
quasars with absorption lines. In this way, we can talk about the 
existence of
yet another large-scale inhomogeneity in the distribution
of matter at scales $Z \sim 3$, which analogous to the first, was 
named SA2.
\par
To observable parts in the Universe the inhomogeneity picture in the 
distribution of
matter also outlines the existence of a gigantic void (SV) at 
distances, estimated
by us to be within $Z \sim 2-3$.
In a very definitive sense, the SA and SV form a gravitational 
dipole.
The direction, defined by the axis of this dipole, is quite close 
to the
``Axis of the Universe'', which in turn is defined by the orientation
of radiogalaxies (Amirkhanjan, 1994).
\par
We assume that the main reason why we observe structures on the 
topographies is
because of effects related to the ``anomalous'' amplification or 
absorption of
quasar radiation during their path through large scale matter 
non-uniformities;
and not because of effects
that arise due to random fluctuations or selections. The basis for 
such an assumption
are the statistical calculations described in section 4.
Besides, we carried out a series of tests to estimate the possible 
extra effects,
connected with very poor statistics in the galactic equator region. 
In particular,
with the help of a standard algorithm (section 2)
we had calculated the behaviour of the median of the apparent 
stellar magnitude
of quasars in different directions on the celestial sphere.
The median, unlike the mean stellar magnitude, is less sensitive to 
random fluctuations
arising due to poor statistics; and the fact that the structures 
observed on it's topography agree with the behaviour of the 
extremums
on the lensing topography (Fig.1) speak in favor of a statistical 
significance
of these extremums.
However, we cannot totally exclude selection effects. Results of yet 
another test,
related to the analysis of the depth of selection of catalog objects,
show that
there exists some correlation between that region on the celestial 
sphere where
the limiting apparent magnitude is minimum ( i.e. possibly only 
bright objects
were accounted for ) and the
direction towards SA2. With the presence of such an observable 
selection
it could be possible to also explain the ``effective'' amplification 
in the mean
brightness of quasars along a given direction and the presence of a 
maximum in
the absorption bands of their spectra.
\par
Effects of absorption and lensing of quasar radiation by massive 
objects
that form the medium through which this radiation traverses, seem to 
be a real way
to detect and analyze clustered structures of dark matter at scales 
$Z \sim 1$.
In some cases, one can get more information about the
presence of such clusters from the chemical analysis of quasar 
emission spectra.
In the work by Gnedin \& Ostriker (1997), for example, it says
that the process of heirarchical clustering of neutral gas in the 
early
Universe could have possibly led to the formation of clusters of 
massive stars.
An intensive burning of light elements and a generation of heavier 
elements in
these stars leads to a strong ``enrichment'' of
nearer regions with heavy elements. So, the presence of an increased
composition of heavy elements in the observable region may be proof 
of occurrences
of strong matter-clustering processes there.
\par
Using this method, we carried out a comparitive analysis of spectral 
lines of quasars
from the Hewitt \& Burbidge catalog (1993) in different parts of the 
celestial
sphere. The results obtained tell us about a slight excess over the 
background
of the relative composition of heavy
elements in regions adjoining the SA and SA2. Even though these are 
based on poor
statistics, yet in conjunction with other results, they
confirm our conclusions about the existence of gigantic clusters of 
matter at scales
$Z \sim 1$. This is in agreement with the opinion of
Sylos Labini (1996), that the known clusters and galactic 
superclusters are not
the most large scaled structures.

\begin{acknowledgements}
We are very grateful to T.A.Agekian, V.A.Antonov and Y.V.Baryshev 
for providing
useful discussion of the results; to the anonymous referee of
the initial version of this paper for his critical remarks, 
especially for his 
suggestion to switch to the galactic system of coordinates, which 
resulted in our
finding of clump No.2; and to Hewitt \& Burbidge
for the QSO catalog which they delivered.
\end{acknowledgements}

\newpage

{\large {\bf
			Captions to Figures

}}

\vspace{1.5cm}

Figure 1. Topography of the lensing parameter $K_i$ in the 
Mercatorial
projection of the celestial sphere, for four different intervals
of $Z_e$. Radius of the standard cell, $R = 40^\circ$. Number of cell
s,
$N_{knots} = 150$.
\newline
$\bf A)$  $Z_e \epsilon [0.5 ; 2.5]$\newline
$\bf B)$  $Z_e \epsilon [1 ; 3]$\newline
$\bf C)$  $Z_e \epsilon [1.5 ; 3.5]$\newline
$\bf D)$  $Z_e \epsilon [2 ; 4.5]$

\vspace{1.5cm}

Figure 2. Topography of:\\
$\bf A)$ The relative density of absorbers, and\\
$\bf B)$ The lensing parameter $K_i(l,b)$\\
both, in the Mercatorial projection of the celestial sphere in the in
terval $Z_e \epsilon [1.5 ; 3.5]$. Radius of the standard cell, $R =
40^\circ$. Number of cells, $N_{knots} = 150$.

\newpage

$\bf Table I.$
Results of the probability estimate of random occurrence
of a spot with a minimum $K_i$ on the lensing topography in the
direction $l_1 = (250^\circ \pm 20^\circ)$; $b_1 = (-5^\circ \pm
20^\circ)$, for four intervals of $Z_e$. Radius of the standard 
cell, $R = 40^\circ$. Number of cells, $N_{knots} = 108$. Number of throws or iterations,
$Q = 10^4$.
Explanation is given in the text.

\vspace{0.5cm}

\bf I.a)  $Z_e \epsilon [0.5 ; 2.5]$
\begin{center}
\begin{tabular}[t]{|c|c|c|c|}
\hline 
$\bf K_{cr}$ & 0.963 & 0.968 & 0.972 \\
$\bf m$ & 5 & 6 & 8 \\ 
$\bf n$ & 6 & 7 & 13 \\
$\bf p\%$ & 0.05 & 0.09 & 1.8 \\
\hline
\end{tabular}
\end{center}

\vspace{0.5cm}

\bf I.b)  $Z_e \epsilon [1; 3]$
\begin{center}
\begin{tabular}[t]{|c|c|c|c|}
\hline 
$\bf K_{cr}$ & 0.968 & 0.97 & 0.972 \\
$\bf m$ & 5 & 6 & 7 \\ 
$\bf n$ & 6 & 8 & 11 \\
$\bf p\%$ & 0.05 & 0.2 & 0.6 \\
\hline
\end{tabular}                                    
\end{center}

\vspace{0.5cm}

\bf I.c)  $Z_e \epsilon [1.5 ; 3.5]$
\begin{center}
\begin{tabular}[t]{|c|c|c|c|}
\hline
$\bf K_{cr}$ & 0.963 & 0.967 & 0.97 \\          
$\bf m$ & 2 & 2 & 2 \\                           
$\bf n$ & 3 & 6 & 9 \\                          
$\bf p\%$ & 8 & 42 & 76 \\                 
\hline                                           
\end{tabular}              
\end{center}

\vspace{0.5cm}

\bf I.d)  $Z_e \epsilon [2 ; 4.5]$
\begin{center}
\begin{tabular}[t]{|c|c|c|c|}
\hline
$\bf K_{cr}$ & 0.966 & 0.97 & 0.972 \\          
$\bf m$ & 5 & 6 & 9 \\                           
$\bf n$ & 6 & 8 & 11 \\                          
$\bf p\%$ & 0.05 & 0.2 & 0.7 \\                 
\hline                                           
\end{tabular}              
\end{center}

\newpage

$\bf Table II.$
Results of the probability estimate of random occurrence
of a spot with a minimum $K_i$ on the lensing topography in the
direction $l_2 = (120^\circ \pm 20^\circ)$; $b_2 = (5^\circ \pm
20^\circ)$, for four intervals of $Z_e$. Radius of the standard
cell, $R = 40^\circ$. Number of cells, $N_{knots} = 108$. Number of t
hrows or iterations,
$Q = 10^4$.
Explanation is given in the text.

\vspace{0.5cm}

\bf II.a)  $Z_e \epsilon [0.5 ; 2.5]$
\begin{center}
\begin{tabular}[t]{|c|c|c|c|}
\hline
$\bf K_{cr}$ & 0.971 & 0.973 & 0.975 \\
$\bf m$ & 5 & 8 & 9 \\
$\bf n$ & 12 & 17 & 19 \\
$\bf p\%$ & 2.7 & 6.7 & 14 \\
\hline
\end{tabular}
\end{center}

\vspace{0.5cm}

\bf II.b)  $Z_e \epsilon [1; 3]$
\begin{center}
\begin{tabular}[t]{|c|c|c|c|}
\hline
$\bf K_{cr}$ & 0.972 & 0.974 & 0.976 \\
$\bf m$ & 4 & 7 & 8 \\
$\bf n$ & 11 & 15 & 18 \\
$\bf p\%$ & 6.5 & 3.8 & 8.7 \\
\hline
\end{tabular}
\end{center}

\vspace{0.5cm}

\bf II.c)  $Z_e \epsilon [1.5 ; 3.5]$
\begin{center}
\begin{tabular}[t]{|c|c|c|c|}
\hline
$\bf K_{cr}$ & 0.965 & 0.97 & 0.974 \\
$\bf m$ & 3 & 4 & 7 \\
$\bf n$ & 5 & 9 & 14 \\
$\bf p\%$ & 2.1 & 2.5 & 2.5 \\
\hline
\end{tabular}
\end{center}

\vspace{0.5cm}

\bf I.d)  $Z_e \epsilon [2 ; 4.5]$
\begin{center}
\begin{tabular}[t]{|c|c|c|c|}
\hline
$\bf K_{cr}$ & 0.974 & 0.976 & 0.978 \\
$\bf m$ & 4 & 5 & 8 \\
$\bf n$ & 15 & 16 & 20 \\
$\bf p\%$ & 25 & 12 & 14 \\
\hline
\end{tabular}
\end{center}

\newpage

$\bf Table III.$
Results of the probability estimate of random occurrence
of a spot with a maximum $K_i$ on the lensing topography in the
direction $l_3 = (60^\circ \pm 20^\circ)$; $b_3 = (0^\circ \pm
20^\circ)$, for the interval $Z_e \epsilon [2 ; 4.5]$. Radius of the standard
cell, $R = 40^\circ$. Number of cells, $N_{knots} = 108$. Number of t
hrows or iterations,
$Q = 10^4$.
Explanation is given in the text.

\vspace{0.5cm}

\bf III.  $Z_e \epsilon [2 ; 4.5]$
\begin{center}
\begin{tabular}[t]{|c|c|c|c|}
\hline
$\bf K_{cr}$ & 1.02 & 1.018 & 1.016 \\
$\bf m$ & 3 & 7 & 9 \\
$\bf n$ & 5 & 7 & 9 \\
$\bf p\%$ & 2.1 & 0.08 & 0.2 \\
\hline
\end{tabular}
\end{center}

\newpage

$\bf Table IV.$
Results of the probability estimate of random occurrence
of a spot with a maximum value for the relative density of absorbers
in the
direction $l = (135^\circ \pm 20^\circ)$; $b = (10^\circ \pm
20^\circ)$, for the interval $Z_e \epsilon [1.5 ; 3.5]$. Radius of the standard
cell, $R = 40^\circ$. Number of cells, $N_{knots} = 108$. Number of t
hrows or iterations,
$Q = 10^4$.
Explanation is given in the text.

\vspace{0.5cm}

\bf IV.  $Z_e \epsilon [1.5 ; 3.5]$
\begin{center}
\begin{tabular}[t]{|c|c|c|c|}
\hline
$\bf K_{cr}$ & 3.2 & 2.8 & 2.6 \\
$\bf m$ & 3 & 4 & 6 \\
$\bf n$ & 4 & 8 & 12 \\
$\bf p\%$ & 0.69 & 1.4 & 1.3 \\
\hline
\end{tabular}
\end{center}

\newpage

$\bf Table V.$
Results of the probability estimate of formation of extremal spots
on the lensing topography, occurring due to selection effects.
This was done for four intervals of $Z_e$.
Number of throws or iterations is,
$Q_1 = 10^3$.
Explanation is given in the text.

\bf V.a)
\begin{center}
\begin{tabular}[t]{|c|c|c|}
\hline
$\bf {\Delta Z_e}$ & $\bf R^\circ$ & $\bf{p_1 \%}$ \\
\hline
0.5-2.5 & 30 & 12 \\
0.5-2.5 & 35 & 0.0 \\
0.5-2.5 & 40 & 0.4 \\
\hline
\end{tabular}
\end{center}

\bf V.b)  
\begin{center}
\begin{tabular}[t]{|c|c|c|}
\hline
$\bf {\Delta Z_e}$ & $\bf R^\circ$ & $\bf{p_1 \%}$ \\
\hline
1-3 & 30 & 24 \\
1-3 & 35 & 0.0 \\
1-3 & 40 & 0.9 \\
\hline
\end{tabular}
\end{center}

\bf V.c)  
\begin{center}
\begin{tabular}[t]{|c|c|c|}
\hline
$\bf {\Delta Z_e}$ & $\bf R^\circ$ & $\bf{p_1 \%}$ \\
\hline
1.5-3.5 & 30 & 19 \\
1.5-3.5 & 35 & 0.2 \\
1.5-3.5 & 40 & 0.9 \\
\hline
\end{tabular}
\end{center}

\bf V.d)  
\begin{center}
\begin{tabular}[t]{|c|c|c|}
\hline
$\bf {\Delta Z_e}$ & $\bf R^\circ$ & $\bf{p_1 \%}$ \\
\hline
2-4.5 & 30 & 0.9 \\
2-4.5 & 35 & 0.1 \\
2-4.5 & 40 & 1.3 \\
\hline
\end{tabular}
\end{center}

\newpage

$\bf Table VI.$
Results of the probability estimate of formation of extremal spots
on the lensing topography of the relative density of absorbers,
occurring due to selection effects. This was done for the interval
$Z_e \epsilon [1.5 ; 3.5]$.
The number of
throws or iterations,
$Q = 10^3$.
Explanation is given in the text.

\vspace{0.5cm}

\bf VI.  $Z_e \epsilon [1.5 ; 3.5]$
\begin{center}
\begin{tabular}[t]{|c|c|c|}
\hline
$\bf {\Delta Z_e}$ & $\bf R^\circ$ & $\bf{p_1 \%}$ \\
&& \\
\hline
1.5-3.5 & 30 & 1.7 \\
1.5-3.5 & 35 & 0.2 \\
1.5-3.5 & 40 & 1.2 \\                            
\hline                                           
\end{tabular}
\end{center}

\newpage

\begin{figure}
\centerline{
\epsfig{file=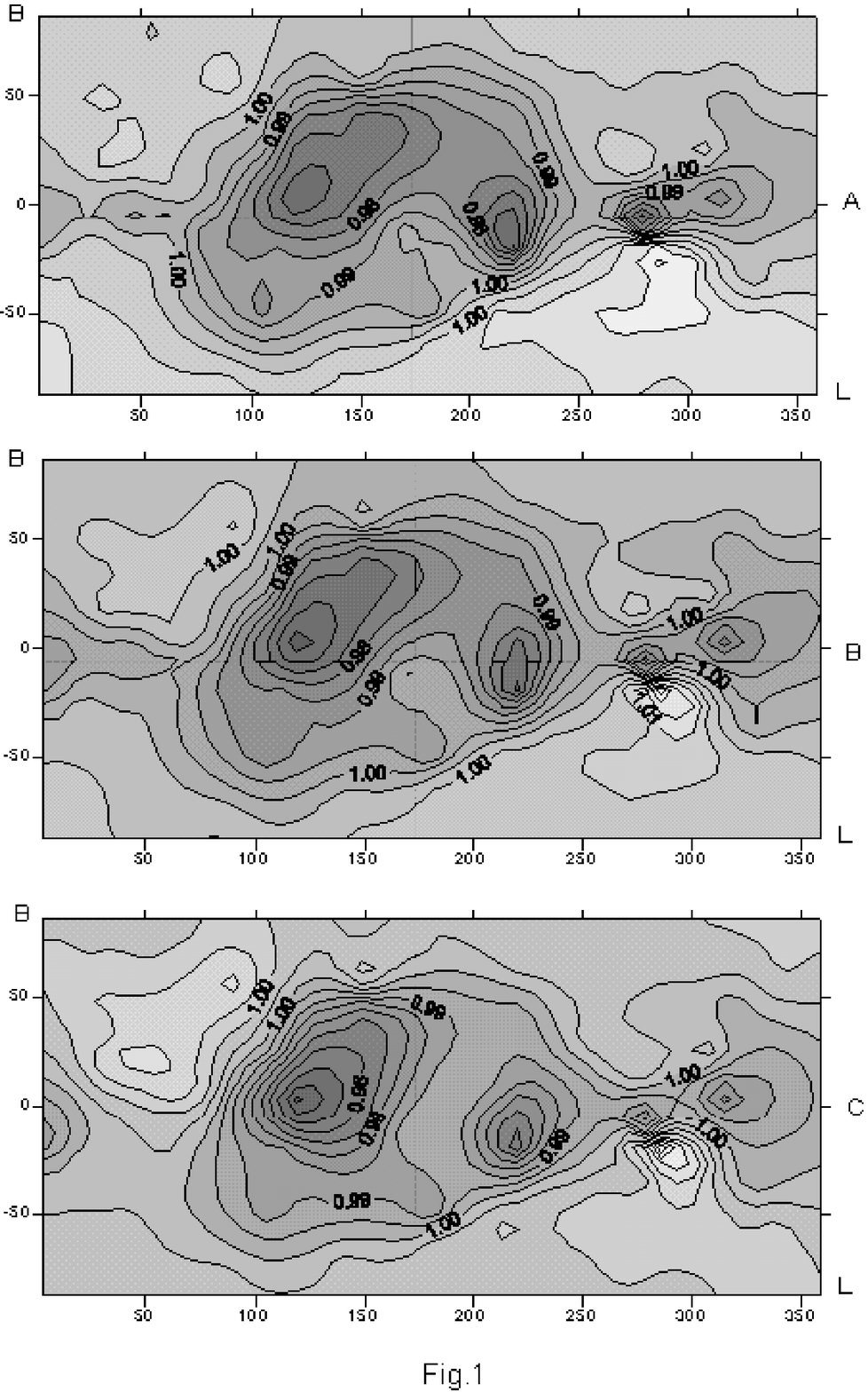,width=28pc}
}\end{figure}

\begin{figure}
\centerline{
\epsfig{file=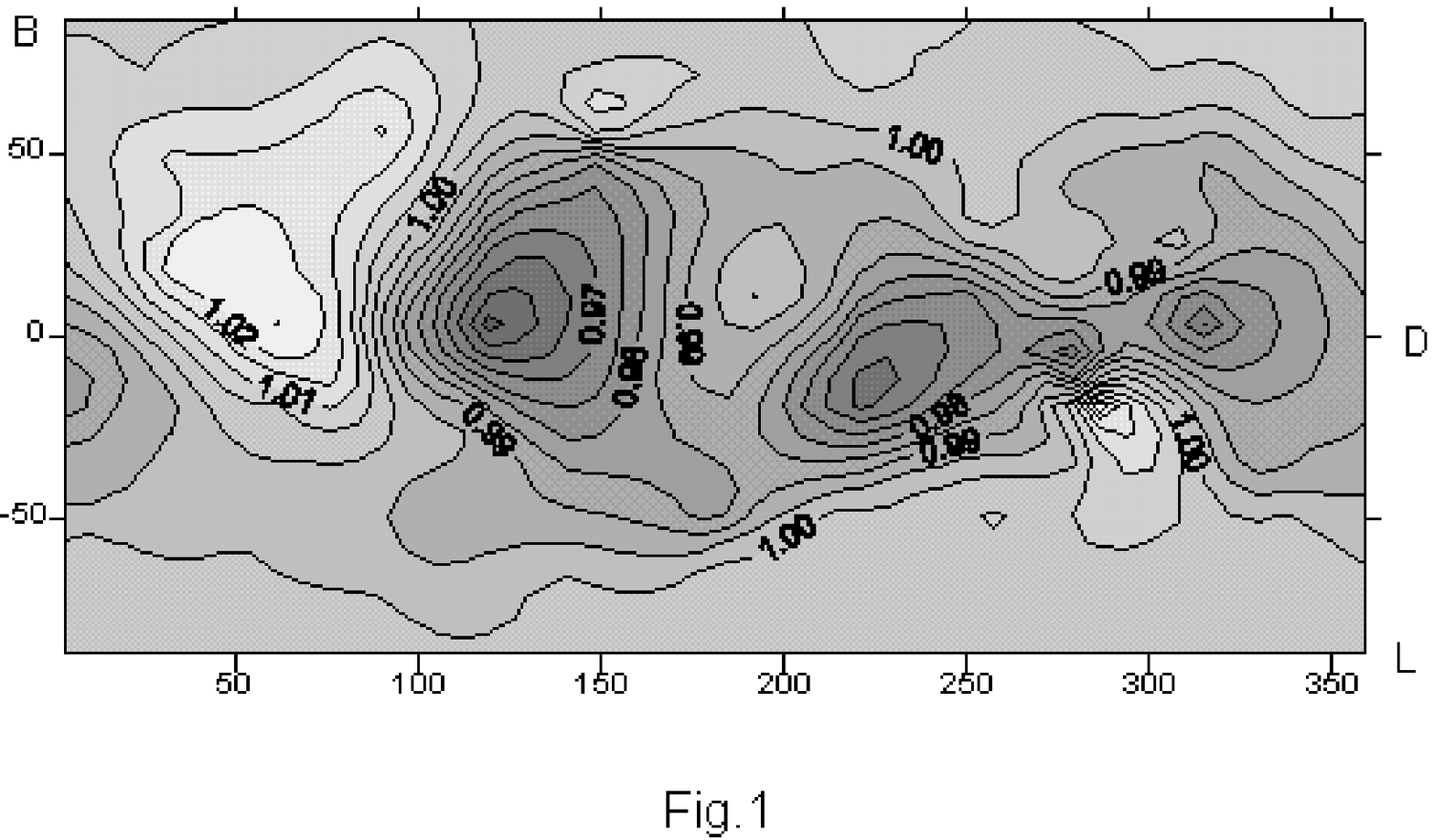}
}
\end{figure}

\begin{figure}
\centerline{
\epsfig{file=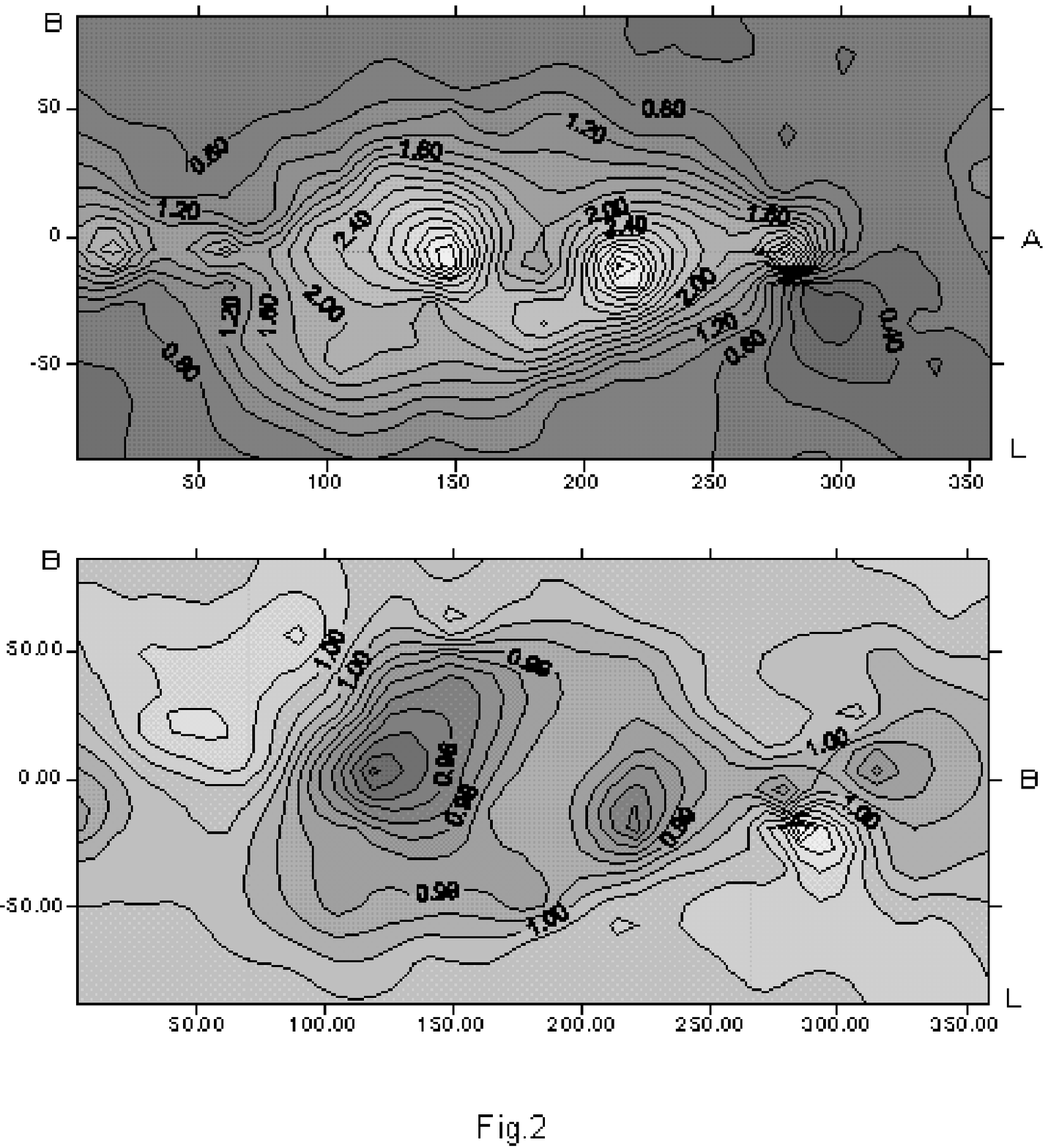}
}
\end{figure}

\end{article}
\end{document}